\documentclass[epsfig,floats,prl,twocolumn,preprintnumbers,floatfix,superscriptaddress]{revtex4-1}
\usepackage{graphicx}
\usepackage[latin2]{inputenc}
\usepackage{amsmath}
\usepackage{amssymb}
\usepackage{bm}
\usepackage{color}

\newcommand{\dldA}{\frac{\text{d}\lambda}{\text{d}\mathcal{A}}\,{\propto}\,(\lambda_{_{\infty}}\!\!\!{-}\lambda)}
\newcommand{\dwdA}{\frac{\text{d}w}{\text{d}\mathcal{A}}\,{\propto}\,w\,(w_{_{\infty}}\!\!\!{-}w)}
\newcommand{\Eqkappa}{\kappa{=}\frac{w_{_{\infty}}}{2w_{_{\text{min}}}{-}w_{_{\infty}}}{-}1}
\newcommand{\Amin}{\mathcal{A}_{\text{min}}}
\newcommand{\wmin}{w(\mathcal{A}{=}\mathcal{A}_{\text{min}}){\equiv}w_{_{\text{min}}}}
\newcommand{\rinv}{{\propto}\,1{/}r}
\newcommand{\tff}{t_\text{f}\,{\sim}\,1{/}f}
\newcommand{\tr}{t_\text{r}}
\newcommand{\trw}{t_\text{r}\,{\sim}\,w^2}
\newcommand{\amaxapp}{f_\text{max}\,{\approx}\,4\,\text{Hz}}

\begin{document}
\title{Striped Patterns in Radially Driven Suspensions with Open Boundaries}
\author{Mahdieh Mohammadi}
\affiliation{Department of Physics $\&$ Optics Research Center, Institute 
for Advanced Studies in Basic Sciences, Zanjan 45137-66731, Iran}
\affiliation{Institute of Physics, Otto von Guericke University Magdeburg, 
39106 Magdeburg, Germany}
\author{Maniya Maleki}
\email{m{\_}maleki@iasbs.ac.ir}
\affiliation{Department of Physics $\&$ Optics Research Center, Institute 
for Advanced Studies in Basic Sciences, Zanjan 45137-66731, Iran}
\author{Adam Wysocki}
\affiliation{Department of Theoretical Physics $\&$ Center for Biophysics, 
Saarland University, 66123 Saarbr\"ucken, Germany}
\author{M. Reza Shaebani}
\email{shaebani@lusi.uni-sb.de}
\affiliation{Department of Theoretical Physics $\&$ Center for Biophysics, 
Saarland University, 66123 Saarbr\"ucken, Germany}

\begin{abstract}
We study the motion of radially driven fluid-immersed particles in a novel Hele-Shaw cell with 
open boundaries. The initially uniform suspension forms a striped pattern within a specific 
range of horizontal oscillation frequencies and for sufficiently large amplitudes. We observe 
that the initial coarsening dynamics of the stripes gradually slows down and the pattern 
reaches a steady state after a few minutes. The distance between the stripes in the steady 
state exhibits an exponentially saturating increase with increased oscillation amplitude or 
frequency. The width of the stripes decreases as a power-law with the frequency while its 
amplitude dependence follows a logistic function. We propose a mechanism--- based on the 
interplay between shear stress, hydrodynamic interactions, and frictional forces--- to link 
the structural characteristics of the stripes to the properties of the oscillatory external 
drive. 
\end{abstract}

\maketitle

Formation of elongated patterns is ubiquitously observed in a wide variety of horizontally 
driven granular systems. Examples range from ripples in sandy deserts and along many coasts 
to striped patterns in colloids, suspensions, dry granular media, and nanoparticle composites 
\cite{Andreotti06,Zhao2016,Reichhardt18,Moosavi14,Klotsa09,Vissers11,Zoueshtiagh08,Reis04,
Mullin00,Sanchez04,Reis02,Roht18,Yuan18,mazzuoli19}. Under horizontal shaking or oscillatory 
excitation, various types of instabilities may arise in dry granular media and suspensions. 
This can lead to, {\it e.g.}, separation of different species \cite{Lozano15}, shear-induced 
segregation \cite{Hill08,Barentin04,Garcimartin17} in environments with a nonuniformly 
distributed shear strain \cite{Moosavi13,Fenistein03}, or, more frequently, stripe formation 
\cite{Sanchez04,Reis02,Reis04,PicaCiamarra05,Mullin00,Pooley04,Fujii12,PicaCiamarra06,Wysocki09,
Krengel13,Mobius14,Roht18}. 

Although identifying the criteria under which the stripes emerge has been the focus of 
many studies, the underlying mechanisms have not been fully understood to date; one cannot 
reliably predict whether stripes emerge in a given granular system under a certain external 
drive. The influential factors on formation and stability of structures in dry granular 
media are frictional contacts \cite{Krengel13,Mobius14,Shaebani08,Goldenberg05} and 
inelastic collisions which lead to effective (long-range) interactions \cite{Zuriguel05,
Shaebani12,Shaebani13}. In colloids and suspensions, hydrodynamic forces and viscous drag 
also play a crucial role \cite{Sanchez04,Klotsa09,Wunenburger02}. While the origin of 
stripe formation in granular mixtures is often attributed to the differences between the 
components \cite{Sanchez04,Reis02,Reis04,PicaCiamarra05,Mullin00,Pooley04,Fujii12,PicaCiamarra06,
Wysocki09}, stripes form even between identical particles \cite{Krengel13,Klotsa09,Moosavi14}. 
The role of lateral walls in stripe formation \cite{Loisel15,Krengel13,Roht18} and whether 
the coarsening of the patterns leads eventually to phase separation \cite{Reis06,Sanchez04,
PihlerPuzovic13,Moosavi14} have also been under debate. 

\begin{figure}[b]
\centering
\includegraphics[width=0.47\textwidth]{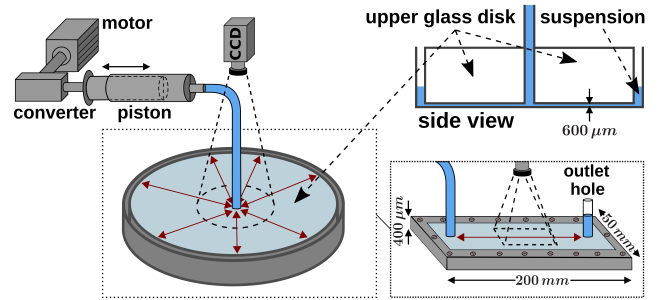}
\caption{Sketch of the experimental setup. The right inset represents the alternative 
rectangular setup. The double-headed red arrows indicate the direction of oscillation.}
\label{Fig:1}
\end{figure}

The detailed structural characteristics of the stripes--- such as the distance between them 
(hereafter called their wavelength $\lambda$) or their width $w$--- are even less understood. 
$w$ was reported to grow with the amplitude $\mathcal{A}$ of the oscillatory drive in 
suspensions \cite{Sanchez04}. Although there is a consensus that the increase of $\mathcal{A}$ 
enhances $\lambda$ in ripples and stripes, different $\lambda{-}\mathcal{A}$ dependencies 
have been proposed so far: While a linear increase of $\lambda$ with $\mathcal{A}$ is often 
reported--- {\it e.g.}\ in granular mixtures \cite{Sanchez04,PicaCiamarra05}, sand ripples 
under water \cite{Stegner99}, suspensions \cite{Roht18,Roht17}, and even for immersed particles 
in a capillary tube \cite{Zoueshtiagh08}---, other dependencies of $\lambda$ on $\mathcal{A}$--- 
such as a power-law relation with an exponent smaller than one in suspensions \cite{Wunenburger02} 
or a faster than linear increase in sand ripples \cite{Ivanova02}--- were also reported. 
For the frequency dependence of $\lambda$, various contradicting relations can be found 
in the literature: A power-law decay of $\lambda$ with $f$ (though with different exponents) 
has been reported for sand ripples \cite{Rousseaux04}, heavy beads immersed in water 
\cite{Wunenburger02}, and binary granular mixtures \cite{Pooley04}. In contrast, an 
increase of the form $\lambda\,{\sim}\,f^2$ in granular mixtures \cite{Sanchez04}, a 
growth faster than $f^2$ in sand ripples \cite{Ivanova02}, and an increase and gradual 
saturation in granular mixtures \cite{PicaCiamarra05} have been reported as well. $\lambda$ 
was even found to be independent of the frequency of the oscillatory drive in suspensions 
\cite{Roht18,Roht17}. In the absence of a unified theoretical framework, it is generally not 
clear how the interplay between the external drive and the nature of interparticle 
and particle-wall interactions dictates $w$ and $\lambda$. A nonuniform drive--- as, 
{\it e.g.}, exerted by propagating waves in curved coasts--- makes the problem even 
more complicated.

Here, we introduce a novel Hele-Shaw cell in which a suspension of a single type of 
particles is radially driven under open and periodic boundary conditions in the radial 
and angular directions, respectively (Fig.\,\ref{Fig:1}). The new setup mimics the 
boundary conditions in curved coasts and also enables us to study a broad range of 
horizontal oscillation amplitudes simultaneously. The emergence of curved striped 
patterns in this setup proves that the lateral walls perpendicular to the driving 
direction are not required for stripe formation. We characterize the structure by 
analyzing the radial-dependence of $w$ and $\lambda$ and show that the growth of 
both quantities with $\mathcal{A}$ ultimately saturates at high driving amplitudes. 
We propose a self-amplification mechanism for the evolution of $w$ which results 
in a logistic growth versus $\mathcal{A}$, as observed experimentally. By considering 
the interplay between streaming flow fields, shear stress exerted by the nonuniformly 
driven fluid, and interparticle and particle-wall frictional forces, we propose a 
mechanism which eventually stops the coarsening process and accounts for the observed 
relations between the structural characteristics of the stripes and the properties 
of the oscillatory drive. 

{\it Setup.---} Our experimental system consists of a hollow cylindrical plexiglass 
container with a diameter of $R\,{=}\,300\,\text{mm}$. A concentric thick plexiglass 
disk with a smaller diameter of $285\,\text{mm}$ is placed at a distance of $600\,\mu
\text{m}$ from the bottom plate. An inlet hole is created in the center of the glass 
disk and is connected through a soft silicone tube and a syringe to a mechanical 
converter, which transforms the rotational motion of an AC motor into a sinusoidal 
vibration (Fig.\,\ref{Fig:1}). The periodic motion of the converter is transferred 
to the syringe to induce a back and forth motion of the suspension inside the tube 
and cell. The suspension is radially driven in the spacing between the parallel 
glass surfaces at the bottom and is exposed to the open air at the lateral gap 
between the cylinder and the disk. The oscillation frequency $f$ is controlled by 
an AC speed controller and the amplitude of the piston is held constant at $21\,
\text{mm}$, which results in a radial fluid displacement of, {\it e.g.}, $\mathcal{A}
\,{\simeq}\,10\,\text{mm}$ at the distance $10\,\text{mm}$ from the axis. Besides 
this main setup, a second rectangular setup is also used for complementary experiments; 
see Fig.\,\ref{Fig:1}(inset) and \cite{Moosavi14}.

\begin{figure}[b]
\centering
\includegraphics[width=0.47\textwidth]{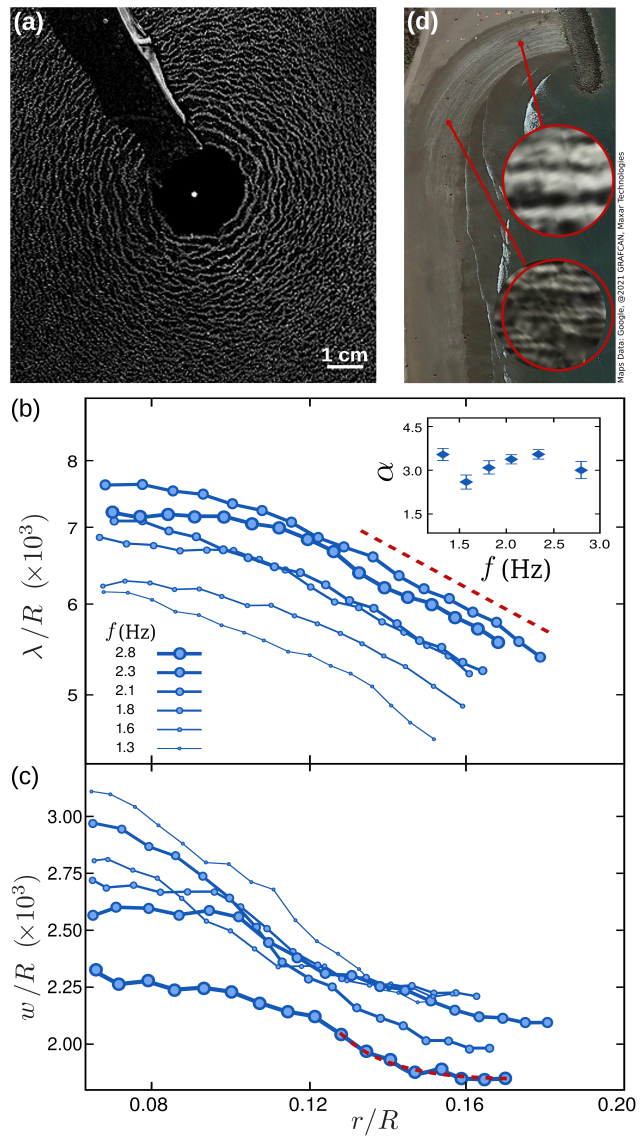}
\caption{(a) Representative image of the steady stripes in a suspension vibrated at 
$f{=}2.1\,\text{Hz}$. (b) $\lambda$, scaled by the radius of curvature $R$ at the 
open boundary, vs $r{/}R$ in log-lin scale. The dashed guide line represents an 
exponential decay. Inset: Slope of the exponential tail vs $f$. (c) $w$ versus 
$r{/}R$. The dashed line shows a $1{/}r$ decay as a guide to the eye. (d) Satellite 
image of a sandy beach near $27^{\circ}45'\text{N},\;15^{\circ}33'\text{W}$ (image 
credit:\ Google Maps), showing a clear difference in the ripple pattern along 
neighboring curved or flat coast lines.}
\label{Fig:2}
\end{figure}

The cell is filled with a granular suspension composed of spherical glass beads 
with density $\rho\,{=}\,2.5\,\text{g}{/}\text{cm}^3$ and diameter in the range 
$\sigma\,{=}\,150{-}200\,\mu\text{m}$ immersed in ethanol (with density 
$\rho\!_{_\text{eth}}{=}\,0.8\,\text{g}{/}\text{cm}^3$ and dynamic viscosity 
$\eta\,{=}\,1.2\,\text{mPa}{\cdot}\text{s}$ at $20\,^\circ\text{C}$) or water. 
We used salt (NaCl) and dishwashing liquid in aqueous suspensions to avoid 
electrostatic effects and for an easier sedimentation of the beads. 
Although similar patterns form in both cases, here we present the results of 
the ethanol suspension. The maximum particle Reynolds number is around $\text{Re}
\,{\approx}\,2$ and the flow generated by the bead is not instantaneous and 
time-reversible. The beads occupy a volume fraction of around $7{\%}$, which 
corresponds to a mean interparticle distance of ${\sim}300\,\mu\text{m}$. We 
use a Canon Rebel T2i camera with a resolution of $110\,\mu\text{m}{/}\text{pixel}$ 
and frame rate of $24\,\text{fps}$ to image the suspension from above. To characterize 
the structural properties of the patterns, the contrast between the solution and 
the beads is increased by placing the setup on a dark plate and illuminating from 
above. In the circular setup, the widths of successive stripes and the spacing 
between them along differently oriented radii are measured and a radial coordinate 
$r$ is assigned to the center of each recorded stripe width or spacing between two 
neighboring stripes. Next, by binning the radial distance, the results belonging 
to the same bin are averaged over all orientations to obtain $w(r)$ and $\lambda(r)$. 
In the rectangular setup, the measurement is performed along parallel lines oriented 
along the oscillation direction and the mean $w$ and $\lambda$ are calculated over 
the entire data.

{\it Formation of patterns.---} The initially well-mixed suspension becomes unstable 
within a few seconds and subsequently elongated curved structures emerge for a 
sufficiently large amplitude of the piston (practically$\,{>}\,5\,\text{mm}$) 
and within the frequency range $1{-}3\,\text{Hz}$, for which the driving strength 
$\Gamma{\propto}f^2$ is neither too weak (no pattern) nor too strong (vortex 
formation). We do not aim here to precisely identify the stripe formation subdomain 
in the $f{-}\mathcal{A}$ space. The pattern gradually coarsens but reaches a 
dynamical saturation after a few minutes, where the overall shape of the stripes 
becomes time invariant \cite{Moosavi14}. Notably, we have observed striped patterns 
even in a setup where the distance between the plates is slightly larger than 
the size of one single bead. Also we would like to emphasize that no visual sign 
of vortices or turbulent flow has been observed since the analyzed region 
($10\,\text{mm}\,{<}\,r\,{<}\,60\,\text{mm}$) is far away from both boundaries. 
Indeed, vortices may form near the inlet pipe. The size $L$ of the affected 
region is given by $L{/}D\,{=}\,0.39\,Re_{i}^{1/3}(H{/}D)^{4/5}$, where $D\,{=}\,
6\,\text{mm}$ is the inlet diameter, $H\,{=}\,0.6\,\text{mm}$ the gap height 
between the two plates, and $Re_{i}\,{\approx}\,400$ the maximal inlet flow Reynolds 
number \cite{Stergiou22}. Accordingly, the vortex length cannot be larger than 
$L\,{\approx}\,3\,\text{mm}$ in our experiment. Furthermore, $Re_{i}$ as well 
as the flow Reynolds number between the plates ($Re\,{\approx}\,6$) are much smaller 
than the critical Reynolds number $Re_c\,{\approx}\,2040$ for the onset of 
turbulence in pipe flows \cite{Avila11}. It can be shown that the velocity 
profile between the plates in our setup is a quasi-parabolic profile in the 
vertical dimension and there are small deviations from a time-periodic plane 
Poiseuille flow. Nevertheless, the laminar radial flow is preserved and the 
system is far away from the high Reynolds number and turbulent flow regime.

Figure\,\ref{Fig:2}(a) shows a typical striped pattern in the steady state. Both 
the width $w$ of the stripes and spacing $\lambda$ between them grow as approaching 
the center of the cell, until they ultimately saturate; see Fig.\,\ref{Fig:2}(b),(c). 
The tail of $\lambda$ decays almost exponentially with the radial coordinate $r$, 
{\it i.e.}\ $\lambda\,{\sim}\,\text{exp}[-\alpha\,\frac{r}{R}]$ with a $f$-independent 
decay rate, $\alpha$ as shown in the inset of Fig.\,\ref{Fig:2}(b). The $\lambda{-}r$ 
curves shift to higher wavelengths with increasing $f$ but converge at high frequencies. 
The tail of $w{-}r$ curves decays relatively slower and can be approximated as 
$w\,{\sim}\,1{/}r$. With a further reduction of $\lambda$ towards the open boundary, 
the pattern gradually becomes indistinguishable and the suspension remains uniform 
in the limit $r{\rightarrow}R$. Interestingly, curvature-dependent patterns can be 
also observed in sand ripples formed along curved coasts [see {\it e.g.}\ 
Fig.\,\ref{Fig:2}(d)]: Because of the curvature of the coast line, the incoming plane 
waves are transformed into curved oscillatory wave fronts that fade away towards 
the open boundary at the shore. Sediment transport is a dominant process in ripple 
formation, which plays no role in stripe formation in our well-mixed suspensions; 
nevertheless, the analogies between the two systems ({\it esp.}\ the radial drive 
and open boundaries) and between the emerging patterns suggest that the global 
scales should be irrelevant in the formation of ripples and the mechanism is 
governed by the local scales, without a need for effective long-range interactions 
(as it is demonstrated in the following for the stripes in our driven suspensions). 
A detailed analysis and comparison between the underlying physical mechanisms 
and emerging patterns in these systems will be presented elsewhere.

\begin{figure}[t]
\centering
\includegraphics[width=0.47\textwidth]{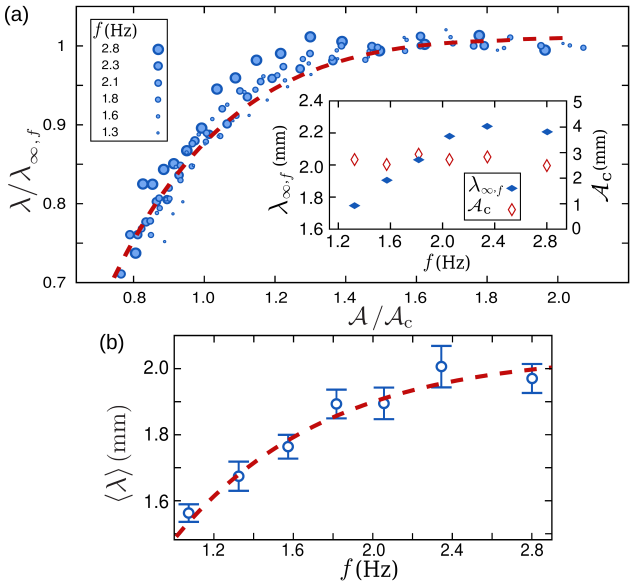}
\caption{(a) $\lambda$, scaled by $\lambda_{_{\infty,f}}$, vs $\mathcal{A}{/}
\mathcal{A}_{\text{c}}$ at different values of $f$. The line is a fit to the 
exponential convergence Eq.\,(\ref{Eq:lambda-A}). Inset: $\lambda_{_{\infty,f}}$ 
and $\mathcal{A}_{\text{c}}$ vs $f$. (b) Averaged wavelength over all oscillation 
amplitudes vs $f$. The line is a fit to Eq.\,(\ref{Eq:lambda-f}).}
\label{Fig:3}
\end{figure}

{\it Dependence of $\lambda$ on $\mathcal{A}$ and $f$.---} To better understand 
the influence of the oscillation amplitude on the distance between the stripes, 
we plot $\lambda$ vs $\mathcal{A}$ (${\propto}\,1{/}r$ since the local displacement 
of the fluid in the circular setup decays with the inverse radial distance from 
the center). $\lambda$ increases with $\mathcal{A}$ with a decreasing rate and 
gradually saturates to a $f$-dependent saturation wavelength $\lambda_{_{\infty,f}}$. 
When scaled by $\lambda_{_{\infty,f}}$, the data for all $f$ values lie on a 
master curve
\begin{equation}
\lambda=\lambda_{_{\infty,f}}\big(1-\text{e}^{-\mathcal{A}{/}\mathcal{A}_{\text{c}}}
\big),
\label{Eq:lambda-A}
\end{equation}
with $\mathcal{A}_{\text{c}}\,{\simeq}\,2.8{\pm}0.2\,\text{mm}$ being the characteristic 
saturation amplitude, {\it i.e.}\ a measure of how fast the maximum possible spacing 
$\lambda_{_{\infty,f}}$ is reached. The fit curve in the main panel of Fig.\,\ref{Fig:3}(a) 
is obtained by fitting the entire data for all frequencies to the exponential convergence 
Eq.\,(\ref{Eq:lambda-A}) using a single fit parameter $\mathcal{A}_{\text{c}}$. Moreover, by 
repeating the fitting procedure for each frequency, we separately obtain $\mathcal{A}_{\text{c}}$ 
for each frequency. The results presented in the inset of Fig.\,\ref{Fig:3}(a) 
show that $\mathcal{A}_{\text{c}}$ remains nearly unchanged with respect to $f$. 
$\lambda_{_{\infty,f}}$ is the maximum possible spacing between the stripes at a given 
frequency. The motility range of the stripes (thus the spacing $\lambda$ between them) 
increases with the fluid motion amplitude until $\lambda$ approaches the upper limit 
$\lambda_{_{\infty,f}}$. At larger driving amplitudes, the stripe motion cannot follow 
the oscillatory flow anymore. $\lambda_{_{\infty,f}}$ grows with $f$ and eventually 
saturates at higher frequencies. We also average the results over the entire range of 
$r$--- {\it i.e.}\ over all oscillation amplitudes--- at each frequency to clarify the 
overall dependence of $\lambda$ on $f$. Figure\,\ref{Fig:3}(b) shows that the behavior 
of the averaged wavelength $\langle\lambda\rangle$ can be described by an exponential 
convergence
\begin{equation}
\langle\lambda\rangle=\langle\lambda\rangle\!_{_{\infty}}\big(1-\text{e}^{-f{/}
f_\text{c}}\big),
\label{Eq:lambda-f}
\end{equation}
where $\langle\lambda\rangle\!_{_{\infty}}$ is the saturation wavelength and 
$f_\text{c}$ the characteristic frequency. A similar behavior of $\lambda$ vs 
$f$ was numerically observed in dry granular systems \cite{PicaCiamarra05}.

\begin{figure}[t]
\centering
\includegraphics[width=0.47\textwidth]{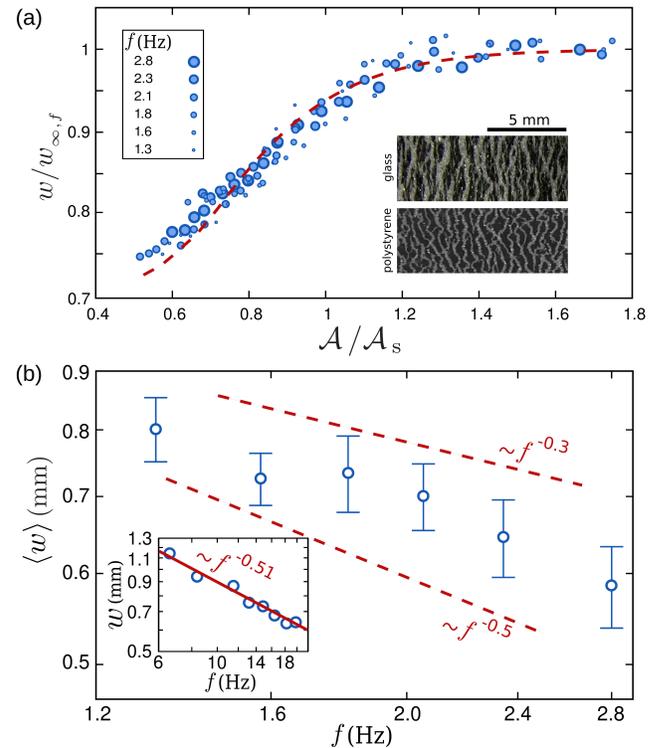}
\caption{(a) $w$, scaled by $w_{_{\infty,f}}$, vs $\mathcal{A}{/}\mathcal{A}_{\text{s}}$ 
at different $f$. The dashed line is the logistic form given by Eq.\,(\ref{Eq:w-A}). 
Insets: Comparison between the stripes formed by glass or polystyrene beads at the 
same driving strength in the rectangular setup. (b) Averaged stripe width over all 
oscillation amplitudes versus $f$ in logarithmic scales. The lines represent power-law 
decays as a guide to the eye. Inset: $w$ vs $f$ for experiments in the rectangular 
setup. The line is a power-law fit, Eq.\,(\ref{Eq:w-f}).}
\label{Fig:4}
\end{figure}

{\it Effects of the oscillatory drive on $w$.---} At a given $f$, $w$ grows 
monotonically with increasing $\mathcal{A}$ towards the center of the setup. The 
growth rate increases with $\mathcal{A}$ far from the center, but then decreases 
and ultimately decays to zero at large amplitudes ({\it i.e.}\ in the vicinity of 
the axis) [Fig.\,\ref{Fig:4}(a)]. The saturation width $w_{_{\infty,f}}$ is thinner 
at higher frequencies. The characteristic saturation amplitude $\mathcal{A}_{\text{s}}$ 
is slightly larger compared to $\mathcal{A}_{\text{c}}$ for $\lambda{-}\mathcal{A}$ 
behavior, but it is similarly independent of $f$.

Next, we average the width over the entire range of $r$ at each frequency. The 
resulting averaged width $\langle w\rangle$ exhibits a weak decay with $f$ 
[Fig.\,\ref{Fig:4}(b)]. Assuming a power-law decay, we obtain an exponent $0.3
{\leq}\,\beta\,{\leq}0.5$ but the $r$-averaged data is not conclusive and it is 
difficult to detect a weak dependency on $f$ because the patterns practically 
emerge over different ranges of $r$ with varying $f$, the frequency window of 
stripe formation in this setup is too narrow, and $w$ is of the order of just 
a few particle diameters. Therefore, we repeat the experiments in the rectangular 
setup shown in Fig.\,\ref{Fig:1}, which enables us to observe the striped 
pattern over a broader range of $f$ and analyze the $w{-}f$ relation more 
precisely. We present the results of experiments with polystyrene beads in 
the inset of Fig.\,\ref{Fig:4}(b). It turns out that $w$ decays as a power-law 
\begin{equation}
w\,{\sim}\,f^{-\beta},
\label{Eq:w-f}
\end{equation}
with $\beta\,{=}\,0.51\,{\pm}\,0.08$. Note that only a region of size $6\,
\text{cm}\times2\,\text{cm}$ (along and perpendicular to the vibration 
direction, respectively) around the center line of the rectangular cell 
is considered ({\it i.e.}\ far from the boundaries), where the flow field 
remains uniform.

{\it Stripe formation mechanism.---} Our proposed mechanism is built upon a few major 
underlying physical processes: (i) A critical driving is necessary to generate 
irreversible collisions between the particles, which induce diffusive motion in 
an otherwise non-Brownian system \cite{Pine05,Corte08,Guasto10,Metzger13,Jeanneret14}; 
it is known that noise is an important driver of phase separation \cite{Bray02}. 
(ii) A closer look at the steady-state structures reveals that the particles are 
in physical contact with each other and also with the bottom surface. Such frictional 
structures are hyperstatic \cite{Shaebani09,Unger05} and resist against perturbations 
\cite{Shaebani07}; the extent of the contact network enhances the stability of the 
structure. (iii) On the contrary, the shear stress $\tau$ induced by the nonuniformly 
driven fluid tends to break the contacts; $\tau$ is stronger at a higher $f$. (iv) 
The oscillatory motion of a particle in the fluid generates a streaming flow field 
around the particle with a penetration depth $\delta\,{\sim}1{/}\sqrt{f}$ \cite{Klotsa09,
Wunenburger02,Klotsa07,Otto08}; this vorticity pattern sucks or ejects the fluid 
towards or away from the particle, which induces repulsive (attractive) interactions 
along (perpendicular to) the vibration direction. The superposition of such hydrodynamic 
interactions was shown to create elongated structures perpendicular to the vibration 
direction \cite{Klotsa09}. 

By vibrating the initially well-mixed suspension, particles collide randomly due to 
slight differences between their actual displacement amplitudes and due to differences 
in the velocity of particles moving at different heights in a Poiseuille flow, which 
leads to the formation of frictional contacts. On the other hand, contacts may break 
due to oscillatory shear forces. At a sufficiently high particle density and for a 
moderate driving strength, the rate of contact formation surpasses the breakage rate, 
which results in the nucleation of small units. The interactions between these units 
through their streaming flow fields cause elongated structures, as explained 
above. The stripes coarsen by random merging of smaller units or individual colliding 
particles. Nevertheless, the dynamics of the structures is limited to the local 
oscillation amplitude of the fluid, thus, the coarsening process is short ranged. 
Upon increasing the stripe width, a weak hydrodynamic repulsion gradually develops 
between the neighboring stripes in the vibration direction, which eventually arranges 
the stripes at distances $\lambda\,{\propto}\,\mathcal{A}$. However, $\lambda$ does 
not continue to grow with $\mathcal{A}$ at an extremely large $\mathcal{A}$ because 
the weak hydrodynamic repulsion cannot push the stripes over long distances. In this 
regime, $\lambda$ and $w$ grow until the contact area between the stripe and the 
bottom plate becomes large enough that the stripe-wall frictional force ultimately 
balances the shear force which tends to slide the stripe over the surface. This 
leads to the saturation distance $\lambda_{_{\infty}}$ between the stripes. The 
empirical observation Eq.\,(\ref{Eq:lambda-A}) implies that the rate of increasing 
$\lambda$ with respect to $\mathcal{A}$ is proportional to the distance from the 
saturation wavelength, {\it i.e.}\ $\dldA$.

Increasing $f$ has a twofold effect: First, it strengthens $\tau$ against the 
interparticle friction. This enhances the contact breakage rate, which reduces 
$w$; second, it also strengthens $\tau$ against the particle-wall friction such 
that the balance between them is reached at a larger $\lambda$, where the stripe 
becomes sufficiently thick. Nevertheless, we observe that $\lambda$ finally 
saturates with $f$. An object of size $w$ can follow the oscillatory fluid only 
if the characteristic time of the fluid motion $\tff$ is larger than the relaxation 
time $\tr$ of the object in the fluid (which scales as $\trw$ in low Reynolds 
number regime \cite{Geng20}). This sets an upper limit on $f$ for a given object 
size, above which the shear cannot efficiently move the object. We obtain $\amaxapp$ 
for a stripe of width $w\,{=}\,3\sigma$, which is consistent with the frequency 
window $1\,\text{Hz}\,{<}\,f\,{<}\,3\,\text{Hz}$ for pattern formation in our 
circular setup (For $f\,{>}\,3\,\text{Hz}$, we observe that no striped pattern 
forms, vortices appear, and the laminar radial flow is not preserved anymore). Also, 
synchronization of two timescales at the steady state results in $w\,{\sim}\,
1{/}\sqrt{f}$, in agreement with the scaling relation Eq.\,(\ref{Eq:w-f}). 

According to the proposed scenario, decreasing the friction coefficient between 
the particles destabilizes the structures against shear forces and, thus, leads 
to thinner stripes. We confirm the validity of this prediction by repeating the 
experiments in the rectangular setup with polystyrene beads which have a smaller 
friction coefficient compared to the glass beads. As shown in the inset of 
Fig.\,\ref{Fig:4}(a) for experiments at $f{=}19.8\,\text{Hz}$, thicker stripes 
form in the case of glass beads. A similar conclusion can be drawn when comparing 
glass and polystyrene data at other frequency values.

The growth of $w$ by random merging of smaller structures follows a self-amplification 
mechanism: A thicker stripe has a larger contact area, thus, a higher chance for 
further merging events. It also contains a wider frictional contact network which 
is more stable and reduces the particle loss rate by the oscillatory shear. However, 
the coarsening process gradually slows down as the number of remaining floating 
objects in the environment reduces. The available resources are bounded to the 
spacing between the stripes, which is controlled by $\mathcal{A}$. Hence, we propose 
that the growth rate of $w$ vs $\mathcal{A}$ is given by $\dwdA$, thus, the amplitude 
dependence of the coarsening process follows a logistic growth form
\begin{equation}
w=\frac{w_{_{\infty}}}{2}\Big(1+\frac{1}{1{+}\kappa\,\text{e}^{-(\mathcal{A}{-}
\mathcal{A}_{\text{min}})/\mathcal{A}_{\text{s}}}}\Big),
\label{Eq:w-A}
\end{equation}
with $\Eqkappa$ and $\Amin$ being the minimum amplitude at which the patterns emerge, 
{\it i.e.}, $\wmin$. Indeed, three of the model parameters can be extracted from the 
experimental data and there remains only a single fit parameter $\mathcal{A}_{\text{s}}$. 
When rescaled to the maximum possible width $w_{_{\infty,f}}$, the asymptotic value 
of the rescaled width $\frac{w}{w_{_{\infty,f}}}$ fluctuates around $1$ with less 
than $3\,\%$ error for all frequencies; see Fig.\,\ref{Fig:4}(a). Thus, we fit the 
entire $w{/}w_{_{\infty}}$ data (i.e.\ for all frequencies) to Eq.\,(\ref{Eq:w-A}) 
for simplicity. The minimum possible stripe width $w_{_{\text{min}}}$ is taken as 
the stripe width at the maximum radial distance above which the pattern is 
indistinguishable (corresponding to $\mathcal{A}_{_{\text{min}}}\,{\simeq}\,1.8{\pm}
0.2$). By fitting the experimental data to the logistic growth form, we obtain 
$\mathcal{A}_{\text{s}}\,{\simeq}\,3.6{\pm}0.2\,\text{mm}$. As shown in 
Fig.\,\ref{Fig:4}(a), Eq.\,(\ref{Eq:w-A}) and the experimental data are in 
remarkable agreement. Also the leading $\mathcal{A}$-dependent term at small 
amplitudes is proportional to $\mathcal{A}$ ({\it i.e.}\ $\rinv$) in agreement 
with Fig.\,\ref{Fig:2}(c).

In conclusion, we have made a step forward in understanding the mechanism of stripe 
formation in vibrated suspensions by considering the structures at the particle 
level in a novel Hele-Shaw cell. Three ingredients dictate the structure of the 
striped pattern: the oscillatory shear that mixes the suspension, the frictional 
contacts that stabilize the structures and prevent the stripes from sliding over 
the surface, and the streaming flow fields around the stripes that induce repulsion 
between the neighboring stripes while promote the merging of smaller units into 
the growing elongated structures. Our proposed mechanism is entirely governed by 
the local scales and explains the trends of the structural characteristics of the 
stripes in terms of the properties of the oscillatory external drive. Understanding 
the emerging hydrodynamic interactions between the structures in an oscillating 
fluid is still a challenging problem which deserves further detailed studies. 
Our results help to better understand the fundamental processes underlying widespread 
striped pattern formation in vibrated granular systems and can open up a new avenue 
for the design of more efficient technologies for separating particles from a fluid flow.

We thank Robabeh Moosavi for her help with Fig.\,\ref{Fig:4}. M.\ Mohammadi 
acknowledges the hospitality of the group of Prof.\ Ralf Stannarius and partial 
support through the Landesstipendium Sachsen-Anhalt and Deutsche Forschungsgemeinschaft 
(DFG) within project STA 425/46-1. Correspondence and requests for materials 
should be addressed to m{\_}maleki@iasbs.ac.ir or shaebani@lusi.uni-sb.de.

\bibliography{Refs-PatternFormation}

\end{document}